\documentclass[aps,prb,preprint,preprintnumbers,showpacs,superscriptaddress,amssymb,amsmath]{revtex4}
\usepackage{latexsym}
\usepackage[dvips]{graphicx}
\usepackage{bm}
\usepackage{epsfig}
\usepackage{dcolumn}

\begin{document}

\title{X-ray cross-correlation analysis and local symmetries of disordered
systems. I. General theory}

\author{M. Altarelli}
\affiliation{European X-ray Free-Electron Laser Facility, Notkestra\ss e 85, D-22607, Hamburg, Germany}
\affiliation{Deutsches Elektronen-Synchrotron DESY, Notkestra\ss e 85, D-22607 Hamburg, Germany}
\author{R.P. Kurta}
\affiliation{Deutsches Elektronen-Synchrotron DESY, Notkestra\ss e 85, D-22607 Hamburg, Germany}
\author{I.A. Vartanyants}
\email[Reference author: ]{ivan.vartaniants@desy.de}
\affiliation{Deutsches Elektronen-Synchrotron DESY, Notkestra\ss e 85, D-22607 Hamburg, Germany}

\date{\today}

\begin{abstract}

In a recent article (P. Wochner \textit{et al.}, PNAS (2009)) x-ray
scattering intensity correlations around a ring, in the speckle diffraction
pattern of a colloidal glass, were shown to display a remarkable $\sim \cos(n\varphi)$
dependence on the angular coordinate $\varphi$ around the ring, with
integer index $n$ depending on the magnitude of the scattering wavevector.
With an analytical derivation that preserves full generality in the
Fraunhofer diffraction limit, we clarify the relationship of this
result to previous x-ray studies of bond-orientation order, and provide
a sound basis to the statement that the angular intensity correlations
deliver information on local bond arrangements in a disordered (or
partially ordered) system.
We present a detailed analysis of
the angular cross-correlation function and show its applicability for studies of wide range of structural properties of disordered systems,
from local structure to spatial correlations between distant structural elements.
\end{abstract}


\maketitle


\section{Introduction}


In a recent experiment \cite{PNAS} by Wochner \textit{et al.}, partially coherent x-rays with a wavelength of $0.154\;\text{nm}$ were used to investigate a colloidal glass composed of PMMA spheres of $117\;\text{nm}$ radius, with particle suspensions of concentration exceeding the glass formation value. The scattering pattern on a 2D detector has the concentric ring structure characteristic of disordered systems, and the speckle appearance resulting from the partial coherence of the undulator x-rays at the European Synchrotron Radiation Facility (ESRF). The authors of Ref.~\cite{PNAS} have introduced the 4-point angular cross-correlation function (CCF) $C_{q}(\Delta)$ defined as
\begin{equation}
C_{q}(\Delta)=\frac{\langle I(q,\varphi)I(q,\varphi+\Delta) \rangle_{\varphi}-\langle I(q,\varphi)\rangle^{2}_\varphi}{\langle I(q,\varphi)\rangle^{2}_\varphi}, \label{Eq:CqWoch}
\end{equation}
where $I(q,\varphi)$ is the scattered intensity, $q$ is the magnitude of the scattering vector $\mathbf{q}$, $\varphi$ is an angular coordinate around the diffraction ring of radius $q$, and
\begin{equation}
\langle F(\varphi) \rangle_{\varphi} =\frac{1}{2\pi} \int^{2\pi}_{0} F(\varphi)\,d\varphi \label{Eq:Faver}
\end {equation}
denotes an angular average around the ring.\footnote{It is to be noted that the trivial angular dependence due to the linear polarization of the incoming synchrotron beam, important at large scattering angles, must be removed from the intensity.}

The remarkable observation by Wochner \textit{et al.}\cite{PNAS} is that, at least for some $q$ values, $C_{q}(\Delta)$ is well approximated by a cosine function of an integer multiple of the angle $\Delta$, i.e., $C_{q}(\Delta)\sim \cos(n\Delta)$; different values of $n$ are observed for different $q$. In particular, the experiment on PMMA spheres, for $q \simeq 0.04\; \text{nm}^{-1}$, showed a very clean cosine behavior with $n=5$. The authors recovered similar behavior from a numerical simulation, assuming that icosahedral clusters are preferentially formed locally, and computing the scattering intensity and its correlations for a cubic lattice of 8 000 such clusters, with random rotational orientation.

In the following, we derive analytical expressions for the Fourier series expansion of the correlation function in the $0\le \Delta <2 \pi$ interval, from which the cosine-like behavior of the angular correlation function is related to the arrangement and orientation of bond angles and interatomic distances in the system in a completely general way. We limit ourself to the Fraunhofer diffraction case here and leave the discussion of the Fresnel diffraction to a forthcoming publication.
One interesting aspect of this phenomenon is that it is essentially two-dimensional in character; in a disordered three-dimensional (3D) system, it appears that among randomly oriented local bond arrangements, the largest effects are expected from local bond arrangements where n-fold symmetry axes are (at least approximately) lined up with the direction of the incident x-rays. This leads us on the one hand to investigate the  relationship to pure two-dimensional (2D) systems: and in fact extremely marked effects, unveiled by previous x-ray studies of bond-orientational order in liquid crystals \cite{Pindak,Gorecka,Chou} (especially hexatic ones), are strongly related to the recent results of Wochner \textit{et al.}. Our aim is to
provide a sound basis to the statement that the angular intensity correlations deliver information on local bond arrangements in a disordered (or partially ordered) system. In the dilute limit (where local entities containing specific bond angles are separated by distances much larger than the bond lengths) the angular correlation function can be explicitly related to a bond-orientational order parameter, which generalizes the order parameter proposed for hexatic liquid crystals\cite{Chaikin} by Bruinsma and Nelson \cite{Bru}.

In this first paper of a series of two, we give a general theoretical treatment of the problem of the x-ray cross-correlation analysis (XCCA) in a partially disordered system. In the second  paper, we will present the results of various simulations
that demonstrate the general findings presented here. This paper is organized in the following way. In the next section a Fourier series analysis of the intensity angular correlations is presented. In the third section a general theoretical treatment of the CCF is given and the expressions for the Fourier coefficients of the CCF's are derived for the case of a kinematical x-ray scattering. In the fourth section the contribution of different terms to the CCF is analyzed. A special treatment is given to dilute and close-packed systems. In the fifth section we consider correlations in 3D systems, when
the effect of the Ewald sphere curvature becomes important. Conditions, at which the angular CCF shows $\cos{(n\Delta)}$ behavior with odd $n$-numbers will be analysed. The paper is completed by the conclusions and outlook section.


\section{Fourier series analysis of the intensity angular correlations}


We generalize the CCF defined in Eq.~(\ref{Eq:CqWoch}) by introducing the intensity correlations
at two different values of the momentum transfer vectors
\footnote{Similar cross-correlation function, which
includes an additional step of averaging over diffraction patterns, was introduced by Saldin \textit{et al.}\cite{Saldin}.} $q_{1}$ and $q_{2}$
\begin{equation}
C_{q_{1},q_{2}}(\Delta) =
\frac{\left\langle I(q_{1},\varphi)I(q_{2},\varphi+\Delta)\right\rangle _{\varphi} - \left\langle I(q_{1},\varphi)\right\rangle _{\varphi}\cdot\left\langle
I(q_{2},\varphi)\right\rangle_{\varphi}}
{\left\langle I(q_{1},\varphi)\right\rangle _{\varphi}\cdot\left\langle
I(q_{2},\varphi)\right\rangle_{\varphi}},
\label{Eq:Cq1q2_1}
\end{equation}
where the averaging over the angle $\varphi$ is defined in Eq.~(\ref{Eq:Faver}). In the next section we will show, that the magnitudes of the scattering vectors $q_{1}$ and $q_{2}$ are, in fact, the values of the perpendicular components
of the 3D scattering vectors $\mathbf{ q}_{1}$ and $\mathbf{ q}_{2}$.
One can readily see that the CCF (\ref{Eq:Cq1q2_1})
can be rewritten in a slightly different form
\begin{equation}
C_{q_{1},q_{2}}(\Delta)=\frac{\langle(I(q_{1},\varphi)-\langle I(q_{1},\varphi)\rangle_{\varphi})\cdot(I(q_{2},\varphi+\Delta)-\langle I(q_{2},\varphi)\rangle_{\varphi})\rangle_{\varphi}}{\left\langle I(q_{1},\varphi)\right\rangle _{\varphi}\cdot\left\langle
I(q_{2},\varphi)\right\rangle_{\varphi}},
\label{Eq:Cq1q2_2}
\end{equation}
which shows that we are dealing with the angular correlation function
of a normalized deviation of the intensity on the diffraction ring.
Let us define this quantity for each value of the momentum transfer vector $q_{j}$ as
\begin{equation}
D_{I}(q_{j},\varphi)=\frac{I(q_{j},\varphi)-\langle I(q_{j},\varphi)\rangle_{\varphi}}{\langle I(q_{j},\varphi)\rangle_{\varphi}},\quad j=1,2
\label{Eq:Dq_1}
\end{equation}
and notice that this function has obviously vanishing angular average.
The measured correlation function (\ref{Eq:Cq1q2_1}) can therefore be written as
\begin{equation}
C_{q_{1},q_{2}}(\Delta)=\langle D_{I}(q_{1},\varphi)D_{I}(q_{2},\varphi+\Delta)\rangle_{\varphi}.
\label{Eq:Cq1q2_3}
\end{equation}
In order to understand what periodicity or symmetry this function
of $\Delta$ may display, let us now proceed to its expansion into Fourier series
in the $(0,2\pi)$ interval

\begin{subequations}
\begin{eqnarray}
&&C_{q_{1},q_{2}}(\Delta)=\sum\limits_{n=-\infty}^{\infty}C_{q_{1},q_{2}}^{n}e^{in\Delta},\label{Eq:Cq1q2_4}\\
&&C_{q_{1},q_{2}}^{n}=\frac{1}{2\pi}\int_{0}^{2\pi}C_{q_{1},q_{2}}(\Delta)e^{-in\Delta}d\Delta.\label{Eq:Cq1q2n_1}
\end{eqnarray}
\end{subequations}
Here $C_{q_{1},q_{2}}^{n}$ is the $n$-th coefficient in the Fourier series expansion of $C_{q_{1},q_{2}}(\Delta)$.
Substituting now the expression (\ref{Eq:Cq1q2_3}) into Eq.~(\ref{Eq:Cq1q2n_1}) and following the usual arguments for the Fourier transforms of convolutions we get\footnote{Here we used the fact that all functions
in the integrand are periodic functions with the period $2\pi$.}

\begin{eqnarray}
C_{q_{1},q_{2}}^{n}=D^{n \ast}_{I}(q_{1})D^{n}_{I}(q_{2}),
\label{Eq:Cq1q2n_2}
\end{eqnarray}
where $D^{n}_{I}(q_{j})$ are the Fourier coefficients of a normalized deviation of the intensity.
One can see, that in
order to calculate the Fourier coefficients of $C_{q_{1},q_{2}}(\Delta)$, one
may first calculate those of $D_{I}(q_{j},\varphi)$, i.e., $D^{n}_{I}(q_{j})$ and
then take a product according to Eq.~(\ref{Eq:Cq1q2n_2}).
Note, that the definition (\ref{Eq:Dq_1}) of the normalized deviation $D_{I}(q_{j},\varphi)$ implies that $D^{0}_{I}(q_{j})=0$.  Since scattered intensities
are always real quantities, it is also easy to show that $D_{I}^{-n}(q_{j})=D_{I}^{n \ast}(q_{j})$
and, therefore, $C_{q_{1},q_{2}}^{-n}=C_{q_{1},q_{2}}^{n \ast}$.
According to these symmetry conditions
Eq.~(\ref{Eq:Cq1q2_4}) can be represented in the following form
\begin{eqnarray}
&&C_{q_{1},q_{2}}(\Delta)=2\sum\limits_{n=1}^{\infty}\text{Re}\left[C_{q_{1},q_{2}}^{n}e^{in\Delta}\right]
=2\sum\limits_{n=1}^{\infty}\mid C_{q_{1},q_{2}}^{n} \mid \cdot \cos(n\Delta+\gamma_{n}),\label{Eq:Cq1q2_5}\\
&&\gamma_{n}=\text{arg}(C_{q_{1},q_{2}}^{n}),\nonumber
\end{eqnarray}
where the summation is performed over the positive integer numbers $n$.

In the particular case, when $q_{1}=q_{2}=q$, Eqs.~(\ref{Eq:Cq1q2n_2}) and (\ref{Eq:Cq1q2_5})
reduce to
\begin{subequations}
\begin{eqnarray}
&&C_{q}(\Delta)=2\sum\limits_{n=1}^{\infty} C_{q}^{n}\cos(n\Delta),
\label{Eq:Cq1q2_5q}\\
&&C_{q}^{n}=\mid D_{I}^{n}(q)\mid^{2},\quad C_{q}^{n}\geq 0.
\label{Eq:Cq1q2n_2q}
\end{eqnarray}
\end{subequations}

The general analysis of CCF's presented in this section and particularly Eqs.~(\ref{Eq:Cq1q2_5q}, \ref{Eq:Cq1q2n_2q}) explain a single cosine behavior of CCF calculated from experimental data in Ref.\cite{PNAS}. Clearly, a strong single cosine dependence of $C_{q}(\Delta)$ can be observed only for those values of $q$, at which one of the Fourier coefficients $C_{q}^{n}$ significantly dominates over all others.
In the following sections we will show how such coefficients can be related to the structure and symmetry of the system.

It is to be noted that Eqs.~(\ref{Eq:Cq1q2_5q}, \ref{Eq:Cq1q2n_2q}) also imply that the Fourier analysis
of the CCF (\ref{Eq:CqWoch}) investigated by Wochner \textit{et
al.}\cite{PNAS} does not really contain additional information with respect to the Fourier
analysis of the $\varphi$ dependence of the intensity. Examples \cite{Pindak, Gorecka, Chou} of analysis of the periodicity
in the angular dependence of the intensity can be found in studies
of hexatic liquid crystal phases \cite{Chaikin}, performed with incoherent
sources.

In the next section, we present detailed derivations of the CCF, based on the kinematical x-ray scattering theory.


\section{General theoretical treatment of the cross-correlation
function}

\begin{figure*}[!htbp]
\centering
\includegraphics[width=0.9\textwidth]{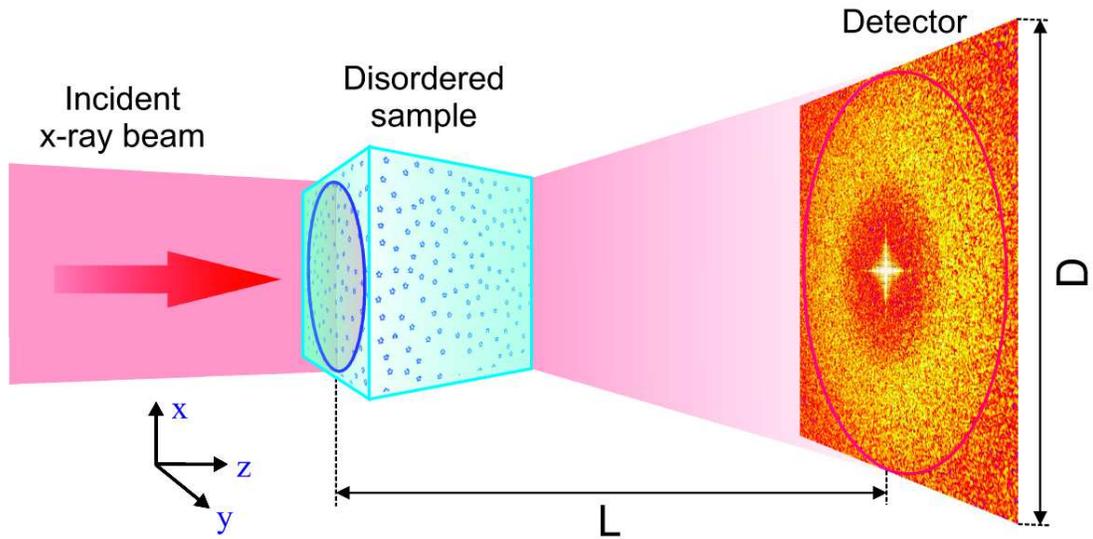}
\caption{\label{Fig:ExpGeometr}(Color online) Geometry of the diffraction experiment.
A coherent x-ray beam illuminates a disordered sample and produces a speckle diffraction pattern on a detector.
The speckle features are defined by the finite size of the beam, or the finite size of the sample and its
microscopic configuration. The direction of the incident beam is defined along the $z$ axis of the coordinate system.}
\end{figure*}
\begin{figure*}[!htbp]
\centering
\includegraphics[width=0.5\textwidth]{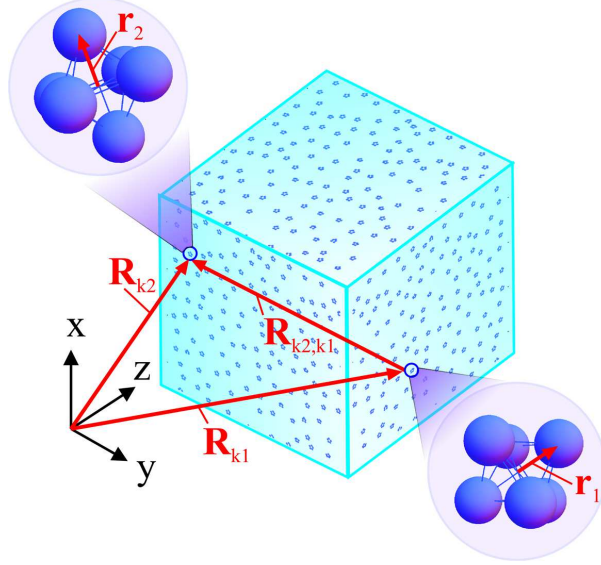}
\caption{\label{Fig:Sample}(Color online) Disordered sample consisting of clusters randomly oriented and
distributed in 3D space. The radius vector $\mathbf{ R}_{k_{2},k_{1}}=\mathbf{ R}_{k_{2}}-\mathbf{ R}_{k_{1}}$
connects the centers $\mathbf{ R}_{k_{1}}$ and $\mathbf{ R}_{k_{2}}$ of two different clusters $k_{1}$ and  $k_{2}$,
the vectors  $\mathbf{ r}_{1}$ and $\mathbf{ r}_{2}$ define the positions of particles inside each cluster
with the origin of a local coordinate system in each cluster $k_{i}$ positioned at its center.
}
\end{figure*}

We start our discussion with a simple scattering geometry depicted
in Fig.~\ref{Fig:ExpGeometr}. A coherent x-ray beam scatters on the disordered sample
and creates a speckle pattern on the detector in the far-field regime.
As a general model system we assume a 3D
sample consisting of identical 3D local structures (LS) of arbitrary shape,
random orientation and position in 3D space (Fig.~\ref{Fig:Sample}).
Such a model includes a variety of systems, i.e., clusters or molecules in the gas phase, LS's formed
in colloidal systems (similar to Ref.\cite{PNAS}), protein molecules,
viruses or complex biological systems in solution.

The coherent x-ray scattering amplitude $A(\mathbf{ q})$ from such a sample
can be described in the first Born approximation (or kinematical scattering)
as
\begin{equation}
A(\mathbf{ q})=\int\rho(\mathbf{ r})e^{i\mathbf{ q\cdot r}}d\mathbf{ r},\label{Eq:Aq1}
\end{equation}
where $\rho(\mathbf{ r})$ is a total electron density of the system
\footnote{In Eq.~(\ref{Eq:Aq1}) we have also tacitly
assumed an infinite illumination region that also means an infinitely small speckle size.
In practice, due to finite size effects (a finite size of the coherent beam, a finite
size of the coherent area in the partially coherent beam, or a finite size of a sample) the size of a speckle is
finite and is of the order $\Delta q \sim 1/d$, where $d$ is a typical length at the sample position.
However, these finite size effects will not influence our further general treatment of the CCF's.}.
For disordered systems under consideration this electron density can
be written in the following form
\begin{equation}
\rho(\mathbf{ r})=\sum\limits _{k=1}^{N}\rho_{k}(\mathbf{ r-R}_{k}),\label{Eq:Ro_1}
\end{equation}
where $\rho_{k}(\mathbf{ r})$ is an electron density of the $k$-th LS at
the position $\mathbf{ R}_{k}$ (see Fig.~\ref{Fig:Sample}) and the summation is performed
over all $N$ LS's. Substituting Eq.~(\ref{Eq:Ro_1}) into Eq.~(\ref{Eq:Aq1}) we
obtain for the total scattered amplitude
\begin{equation}
A(\mathbf{ q})=\sum\limits _{k=1}^{N}e^{i\mathbf{ q\cdot R}_{k}}A_{k}(\mathbf{ q}),\label{Eq:Aq2}
\end{equation}
where $A_{k}(\mathbf{ q})$ is the amplitude scattered by one LS,
\begin{equation}
A_{k}(\mathbf{ q})=\int\rho_{k}(\mathbf{ r})e^{i\mathbf{ q\cdot r}}d\mathbf{ r},\label{Eq:Aq3}
\end{equation}
and the integration is performed over the volume of each LS. 
Eqs.~(\ref{Eq:Aq2}) and (\ref{Eq:Aq3}) express a simple fact that under conditions of coherent
illumination the total scattering amplitude for each value of the wavevector $\mathbf{ q}$
is a coherent sum of the individual amplitudes from each LS modulated with the corresponding
phase term $\text{exp}(i\mathbf{ q R}_{k})$, depending on the position $\mathbf{ R}_{k}$ of each LS.

Using Eq.~(\ref{Eq:Aq2}), we can write the intensity scattered at certain
momentum transfer value $\mathbf{ q}$ as
\begin{eqnarray}
I(\mathbf{ q}) & = & \sum\limits _{k_{1},k_{2}=1}^{N}e^{i\mathbf{ q}\cdot(\mathbf{ R}_{k_{2}}-\mathbf{ R}_{k_{1}})}A_{k_{1}}^{\ast}(\mathbf{ q})A_{k_{2}}(\mathbf{ q})\nonumber \\
 & = & \sum\limits _{k_{1},k_{2}=1}^{N}e^{i\mathbf{ q}\cdot(\mathbf{ R}_{k_{2}}-\mathbf{ R}_{k_{1}})}
 \int\int\rho_{k_{1}}^{\ast}(\mathbf{ r}_{1})\rho_{k_{2}}(\mathbf{ r}_{2})e^{i\mathbf{ q\cdot(r}_{2}-\mathbf{ r}_{1})}d\mathbf{ r}_{1}d\mathbf{ r}_{2}\nonumber \\
 & = & \sum\limits _{k_{1},k_{2}=1}^{N}
 \int\int\rho_{k_{1}}^{\ast}(\mathbf{ r}_{1})\rho_{k_{2}}(\mathbf{ r}_{2})e^{i\mathbf{ q}\cdot \mathbf{ R}_{k_{2},k_{1}}^{21}}d\mathbf{ r}_{1}d\mathbf{ r}_{2}.\label{Intens1}
 \end{eqnarray}
Here, the following notation for the radius vectors connecting two particles $1$ and $2$ in two different clusters
$k_{1}$ and $k_{2}$ was used
\begin{equation}
\mathbf{ R}_{k_{2},k_{1}}^{21}=\mathbf{ R}_{k_{2},k_{1}}+\mathbf{ r}_{21},\label{Eq:Rvec_1}
\end{equation}
where
$\mathbf{ R}_{k_{2},k_{1}}=\mathbf{ R}_{k_{2}}-\mathbf{ R}_{k_{1}}$
is the radius vector connecting different local structures, and
$\mathbf{ r}_{21}=\mathbf{ r}_{2}-\mathbf{ r}_{1}$
is the radius vector connecting subunits inside LS's (see Fig.~\ref{Fig:Sample}).

\begin{figure*}[!htbp]
\centering
\includegraphics[width=0.8\textwidth]{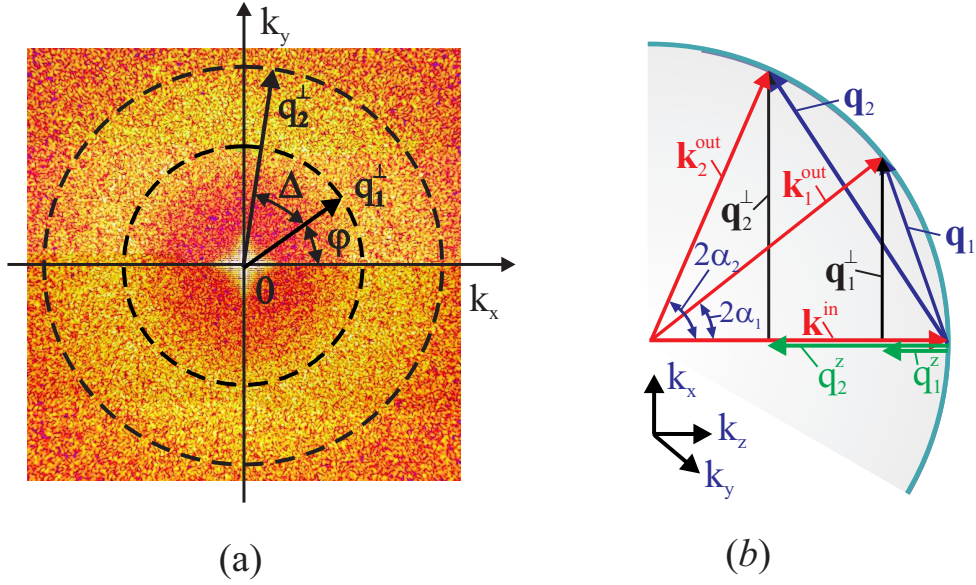}
\caption{\label{Fig:EwaldSphere}(Color online) (a) In the general case, the cross-correlation function can be defined as an angular
average over two intensity rings with different magnitudes of the scattering vectors ${ q}_{1}^{\perp}\neq|{q}_{2}^{\perp}$. The perpendicular components of the scattering vectors $\mathbf{ q}_{1}^{\perp}$ and $\mathbf{ q}_{2}^{\perp}$
are defined in the plane $(k_{x},k_{y})$ in  the polar coordinate system as
$\mathbf{ q}_{1}^{\perp}=(q_{1}^{\perp},\varphi)$ and $\mathbf{ q}_{2}^{\perp}=(q_{2}^{\perp},\varphi+\Delta)$.
(b)  Scattering geometry in the reciprocal space. Here $\mathbf{ k}^{\text{in}}$ is the wavevector of the incident beam
directed along the $z$ axis,
$\mathbf{ k}^{\text{out}}_{1}$ and $\mathbf{ k}^{\text{out}}_{2}$ are the wavevectors of two  scattered waves with the
scattering angles $2\alpha_{1}$ and  $2\alpha_{2}$. The scattering vectors $\mathbf{ q}_{1}=(\mathbf{ q}_{1}^{\perp},q_{1}^{z})$ and $\mathbf{ q}_{2}=(\mathbf{ q}_{2}^{\perp}, q_{2}^{z})$ are decomposed into two components:  $\mathbf{ q}_{i}^{\perp}$ that is perpendicular and $q_{i}^{z}$ that is parallel to the direction of the incident beam.}
\end{figure*}

We decompose now
the scattering vector $\mathbf{ q}=(\mathbf{ q}^{\perp},q^{z})$
into two components: $\mathbf{ q}^{\perp}$ that is perpendicular,
and $q^{z}$ that is parallel to the direction of the incident beam (see Fig.~\ref{Fig:EwaldSphere}).
We define the perpendicular component of the scattering vector $\mathbf{ q}^{\perp}$ in
polar coordinates as $\mathbf{ q}^{\perp}=(q^{\perp},\varphi)$.
We also define the perpendicular
$\mathbf{ R}_{k_{2},k_{1}}^{\perp 21}=\mathbf{ R}_{k_{2},k_{1}}^{\perp}+\mathbf{ r}_{21}^{\perp}$,\
and the $z$-components
$Z_{k_{2},k_{1}}^{21}=Z_{k_{2},k_{1}}+z_{21}$
of the radius vectors introduced in
Eq.~(\ref{Eq:Rvec_1}) (see Figs.~\ref{Fig:ExpGeometr} and \ref{Fig:Sample}).
Using these notations for the vectors we can rewrite Eq.~(\ref{Intens1}) as
\begin{equation}
I(\mathbf{ q})= \sum\limits _{k_{1},k_{2}=1}^{N}e^{-iq^{z}\cdot Z_{k_{2},k_{1}}}
 \int\int\widetilde{\rho}_{k_{1}}^{\ast}(\mathbf{ r}_{1}^{\perp},q^{z})\widetilde{\rho}_{k_{2}}(\mathbf{ r}_{2}^{\perp},q^{z})e^{i\mathbf{ q}^{\perp}\cdot \mathbf{ R}_{k_{2},k_{1}}^{\perp 21}}d\mathbf{ r}^{\perp}_{1}d\mathbf{ r}^{\perp}_{2}.
 \label{Intens2}
 \end{equation}
Here we introduced a modified complex valued electron density function, defined as
\begin{equation}
\widetilde{\rho}_{k_{i}}(\mathbf{ r}_{i}^{\perp},q^{z})=\int\rho_{k_{i}}(\mathbf{ r}_{i}^{\perp},z)e^{-iq^{z}z}dz.
\label{Eq:Ro_2}
\end{equation}

We want to note here that our treatment is quite general and is valid for both cases of wide and small angle scattering.
In the first case, the effect of the Ewald sphere curvature [see Fig.~\ref{Fig:EwaldSphere}(b)], that manifests itself
by the presence of the exponential factors $e^{-iq^{z}\cdot Z_{k_{2},k_{1}}}$ and $e^{-iq^{z}z}$ in Eqs.~(\ref{Intens2}) and (\ref{Eq:Ro_2}), may become important.
This effect could break the scattering symmetry of a diffraction pattern, characteristic for the scattering on a positive valued electron density (Friedel's law) and
may reveal additional symmetries that can be still hidden in the small angle scattering case. As it will be demonstrated in our model simulations this wide angle scattering geometry may become important for a scattering on atomic systems
with local interatomic distances of the order of a few Angstroms.
In the small angle scattering geometry with the scattering angles $2\alpha<<1$
we have for the values of the scattering vectors: $q\simeq2k\alpha(1-\alpha^{2}/6+...)$,
${ q}^{\perp}\simeq2k\alpha(1-2\alpha^{2}/3+...)$; $q^{z}\simeq2k\alpha^{2}(1-\alpha^{2}/3+...)$.
It is well seen from these expressions that the
$q^{z}$ component of the momentum transfer vector is proportional
to the square of the small scattering angle $\alpha$. It means that,
in this situation, the $z$-components of the momentum transfer vectors are
much smaller than their perpendicular components, i.e., $q^{z}<<{ q}^{\perp}$ and can be neglected.
In this limit we have a simplified expression for the intensity (\ref{Intens2})
that does not depend on the $z$-component of the scattering vector $q^{z}$.
For a real valued electron density $\rho_{k_{i}}(\mathbf{ r}_{i})$ the modified electron density function (\ref{Eq:Ro_2})
reduces to a real valued projected electron density of a LS
\begin{equation}
\widetilde{\rho}_{k_{i}}(\mathbf{ r}_{i}^{\perp})=\int\rho_{k_{i}}(\mathbf{ r}_{i}^{\perp},z)dz.
\label{Eq:Ro_3}
\end{equation}
This case of a small angle scattering is typical for scattering on colloidal samples with a typical distances between colloidal particles of few hundred nanometers as in Ref.\cite{PNAS}.

According to Eq.~(\ref{Eq:Cq1q2n_2}), the Fourier coefficients of the CCF are determined by the Fourier coefficients of
the normalized deviation $D_{I}^{n}(q_{j})$. Direct calculations (see Appendix~A for details) give for $D_{I}^{n}(q_{j})$
\begin{equation}
D_{I}^{n}(q_{j})={ I}^{n}({ q}_{j}^{\perp} ,q_{j}^{z})/{ I}^{0}({ q}_{j}^{\perp} ,q_{j}^{z}),\quad n\neq 0, \label{Eq:Dn_1}
\end{equation}
where the Fourier coefficients of the intensity ${I}^{n}({ q}_{j}^{\perp},q_{j}^{z})$
are
\begin{subequations}
\begin{eqnarray}
&&{I}^{n}({ q}_{j}^{\perp},q_{j}^{z}) =(i)^{n} \sum\limits _{k_{1},k_{2}=1}^{N}e^{-iq_{j}^{z}\cdot Z_{k_{2},k_{1}}}{ L}_{k_{1},k_{2}}^{n}({ q}_{j}^{\perp} ,q_{j}^{z}),                                                             \label{Eq:In_1}\\
&&{ L}_{k_{1},k_{2}}^{n}({ q}_{j}^{\perp} ,q_{j}^{z}) =\int\int d\mathbf{ r}_{1}^{\perp}d\mathbf{ r}_{2}^{\perp}\widetilde{\rho}_{k_{1}}^{\ast}(\mathbf{ r}_{1}^{\perp},q_{j}^{z})\widetilde{\rho}_{k_{2}}(\mathbf{ r}_{2}^{\perp},q_{j}^{z})J_{n}({ q}_{j}^{\perp}|\mathbf{ R}_{k_{2},k_{1}}^{\perp 21}|)e^{-in\phi_{\mathbf{ R}_{k_{2},k_{1}}^{\perp 21}}}.
\label{Eq:Ln_k1k2_1}
\end{eqnarray}
\end{subequations}
Here $J_{n}(\rho)$ is the Bessel function of the first kind of integer order $n$, and
$\phi_{\mathbf{ R}_{k_{2},k_{1}}^{\perp 21}}$ is the
azimuthal angle of the perpendicular component of the radius vector
$\mathbf{ R}_{k_{2},k_{1}}^{21}$
defined in Eq.~(\ref{Eq:Rvec_1}) (see Fig.~\ref{Fig:Sample}).

From the derived expressions we can draw the following important conclusions.
According to Eqs.~(\ref{Eq:Cq1q2_4}, \ref{Eq:Cq1q2n_1}, \ref{Eq:Cq1q2n_2}, \ref{Eq:Dn_1}, \ref{Eq:In_1}, \ref{Eq:Ln_k1k2_1}), the initial four-point correlation function $C_{q_{1},q_{2}}(\Delta)$ can be represented by its Fourier series expansion, where each Fourier coefficient is defined by a product of {\it two 2-point} correlation functions of the form (\ref{Eq:In_1}, \ref{Eq:Ln_k1k2_1}), corresponding to two different momentum transfer vectors $\mathbf{ q}_{1}$ and $\mathbf{ q}_{2}$.
The magnitude of $n$-th coefficient is defined by the Fourier coefficients ${ I}^{n}({ q}_{j}^{\perp},q_{j}^{z})$ in Eq.~(\ref{Eq:In_1}), which depend through ${ L}_{k_{1},k_{2}}^{n}({ q}_{j}^{\perp} ,q_{j}^{z})$ on the internal symmetry of LS's as well as on the medium range order of these LS's in the disordered system. We will discuss the structure of these Fourier
coefficients for certain scattering geometries in more detail in the next sections.


\section{CCF decomposition: local structure and interparticle spatial correlations}


In this section we consider more closely the contribution of different terms in the expansion (\ref{Eq:In_1})
to the Fourier coefficients  $C_{q_{1},q_{2}}^{n}$.
We consider here a particular case of a 2D system in a small angle
scattering geometry ($2\alpha<<1$), when we can neglect the $z$-components
of the scattering vectors $q_{1}^{z}$ and $q_{2}^{z}$ . In this case, the
modified electron density\footnote{In this section, we omit the superscript $\perp$, assuming that all vectors are defined in 2D plane.} $\widetilde{\rho}_{k_{i}}(\mathbf{ r}_{i})$ is defined by Eq.~(\ref{Eq:Ro_3}).
The sum in the expression (\ref{Eq:In_1}) for the Fourier coefficients of intensity ${I}^{n}({q}_{j})$ can be split into two parts:
\begin{eqnarray}
{I}^{n}({q}_{j})&\propto&\sum\limits _{k_{1},k_{2}=1}^{N}...=\left[\sum\limits _{k_{1}=k_{2}=k}^{N}...+\sum\limits _{k_{1}\neq k_{2}}^{N}...\right],\label{2.201}
\end{eqnarray}
where the first sum corresponds to the terms with $k_{1}=k_{2}=k$, and the last one to the terms with $k_{1}\neq k_{2}$.

\bigskip
a) \textit{Dilute systems}
\bigskip

It can be shown (see Appendix~B), that for dilute systems, when the average distance $D$ between the clusters is much bigger than the size $d$ of a single cluster, the contribution of the second sum in Eq.~(\ref{2.201}) can become much smaller than that of the first one. In this situation the main contribution to the Fourier coefficients of CCF's will be determined by the first sum in Eq.~(\ref{2.201}) that we will consider in detail below. As soon as for the first term in Eq.~(\ref{2.201}) $k_{1}=k_{2}=k$ and, therefore, $\mathbf{ R}_{k_{2},k_{1}} =0$ and $\mathbf{ R}_{k_{2},k_{1}}^{21}=\mathbf{ r}_{21} $
we have an especially simple expression for the integral $L_{k_{1}=k_{2}=k}^{n}({q}_{j})$ in
Eq.~(\ref{Eq:Ln_k1k2_1})
\begin{eqnarray}
L_{k}^{n}({q}_{j}) & = & \int\int d\mathbf{ r}_{1} d\mathbf{ r}_{2} \widetilde{\rho}_{k}^{\ast}(\mathbf{ r}_{1} )\widetilde{\rho}_{k}(\mathbf{ r}_{2} )J_{n}({q}_{j}|\mathbf{ r}_{21} |)e^{-in\phi_{\mathbf{ r}_{21} }}.\label{2.21a}
\end{eqnarray}
If all LS's have the same internal structure but are oriented and located in space randomly,
the phase $\phi_{\mathbf{ r}_{21} }$ in the exponent of Eq.~(\ref{2.21a}) can be defined as
\begin{equation}
\phi_{\mathbf{ r}_{21} }=\phi_{k}+\phi_{\mathbf{ r}_{21} }^{0},
\label{2phase}
\end{equation}
where $\phi_{k}$ is the rotation angle of the $k$-th LS  with respect to the
fixed angular orientation $\phi_{\mathbf{ r}_{21} }^{0}$ of the LS in the origin of the coordinate system.
In this case, for each LS the integral (\ref{2.21a}) can be expressed in the following form
\begin{eqnarray}
&&L_{k}^{n}({q}_{j})=e^{-in\phi_{k}}{  L}^{n}({q}_{j}).\label{2.22a}
\end{eqnarray}
Here contribution of each LS $k$ is determined by its rotation angle $\phi_{k}$ in the phase and the integral
${  L}^{n}({q}_{j})$ is the same for all LS's
\begin{eqnarray}
&&L^{n}({q}_{j})=\int\int d\mathbf{ r}_{1} d\mathbf{ r}_{2} \widetilde{\rho}^{\ast}(\mathbf{ r}_{1} )\widetilde{\rho}(\mathbf{ r}_{2} )J_{n}({q}_{j}
|\mathbf{ r}_{21} |)e^{-in\phi_{\mathbf{ r}_{21} }^{0}}.\label{LocStr1}
\end{eqnarray}

According to the structure of the integral $L^{n}({q}_{j})$ its value strongly depends on the symmetry of a LS and determines selection rules for the values $n$ of non-zero Fourier coefficients $C_{q_{1},q_{2}}^{n}$. These selection rules can be used for identification of the symmetry of clusters in diluted systems.
For demonstration, we calculate in Appendix~C the integral ${L}^{n}({q}_{j})$ for 2D clusters with the different rotational symmetry (see Fig.~\ref{Fig:SimpleShapes}). For example, for the cluster with 5-fold symmetry [Fig.~\ref{Fig:SimpleShapes}(d)] only $n=10i,\;(i=1,2...)$ will give non-zero contribution to the Fourier coefficients of CCF's. Note, that the Fourier coefficient with $n=5$ is forbidden in this scattering geometry.

In the limit of dilute systems, neglecting the second term in Eq.~(\ref{2.201})
we have for the Fourier coefficients $C_{q_{1},q_{2}}^{n}$ of the CCF
\begin{equation}
C_{q_{1},q_{2}}^{n} \propto {I}^{n*}({q}_{1}) {I}^{n}({q}_{2})\propto \sum\limits_{k=1}^{N}\sum\limits _{k'=1}^{N}...=L^{n\ast}({ q}_{1})L^{n}({ q}_{2})\sum\limits _{k,k'=1}^{N}e^{in\phi_{k',k}},\label{2.23}
\end{equation}
where $\phi_{k',k}=\phi_{k'}-\phi_{k}$.
One can rewrite the sum in Eq.~(\ref{2.23}) in the following form
\begin{equation}
\sum\limits _{k,k'=1}^{N} e^{in\phi_{k',k}}= N^{2} \langle e^{in\phi} \rangle.
\label{2.24}
\end{equation}
Here, the average over all local structure orientations is defined as
\begin{equation}
\langle e^{in\phi} \rangle = \int p(\phi)e^{in\phi}d\phi
\label{2.24a}
\end{equation}
and
\begin{equation}
p(\phi)=1/N^2\sum\limits _{k,k'=1}^{N}\delta(\phi-\phi_{k',k}) \label{Eq:ProbDistr}
\end{equation}
is the probability distribution of angular orientations.
The average $\langle e^{in\phi}\rangle$ is, in fact, a generalization, for $n \neq 6$,
of the bond orientational order parameter,
 introduced for hexatic liquid crystals\cite{Chaikin} by Bruinsma and Nelson \cite{Bru}.

Now, we will consider two different limits for possible orientations of LS's in 2D plane. If all LS's have the same angular orientation, i.e., all $\phi_{k',k}=0$, then the probability distribution function $p(\phi)$ reduces to a
delta function $p(\phi)=\delta(\phi)$. In this case of completely oriented system $\langle e^{in\phi} \rangle =1$.
It means that non-zero values of the Fourier coefficients $C_{q_{1},q_{2}}^{n}$ (\ref{2.23}) will be determined only by the values of the Fourier coefficients ${L}^{n}({q}_{j})$ with the scaling factor proportional to $N^2$.

In another limiting case, when all orientations are uniformly distributed in 2D plane, $p(\phi)=1/(2\pi)$, and
the angular average
\begin{equation}
\langle e^{in\phi} \rangle = \frac{1}{2\pi}\int e^{in\phi}d\phi = \delta_{n,0}.
\label{Eq:UnifDistrib}
\end{equation}
has nonzero value only at $n=0$.
As the Fourier coefficient with $n=0$ is not contributing to the CCF [see Eqs.~(\ref{Eq:Cq1q2_5}, \ref{Eq:Cq1q2_5q})], for a dilute 2D system with random orientations of LS's \textit{all} Fourier coefficients $C_{q_{1},q_{2}}^{n}$ of the angular CCF will be equal to zero. It also means, that in this situation it is not possible to determine the symmetry of LS's from the analysis of the angular CCF. This is similar to the situation
in a small angle x-ray scattering (SAXS), when there is no preferential orientation in the disordered system.

In the case of partial ordering, angular orientations of LS's can be described, for example, by the Gaussian distribution. Such situation may be realized when a disordered system is in an external field (magnetic, electric, \textit{etc}), which drives it towards a more ordered state.
In this case the probability distribution is given by
\begin{equation}
p(\phi)=1/(\sigma\sqrt{2\pi})\text{exp}[-\phi^{2}/(2\sigma^{2})],
 \label{GaussDistrFunc}
\end{equation}
where $\sigma$ is the standard deviation. For this partially ordered state the orientational order parameter
$\langle e^{in\phi} \rangle$ is equal to
\begin{equation}
\langle e^{in\phi} \rangle= \exp{\left(-\frac{1}{2} n^{2}\sigma^{2}\right)}.
 \label{GaussDistr2}
\end{equation}
In this case the number of Fourier coefficients in the CCF is limited.
The strongest contribution to the Fourier coefficients $C_{q_{1},q_{2}}^{n}$ is given by the lowest values of $n$ and is stronger for more ordered systems (that correspond to lower values of $\sigma$).

\bigskip
b) \textit{Close-packed systems}
\bigskip

\begin{figure*}[!htbp]
\centering
\includegraphics[width=0.9\textwidth]{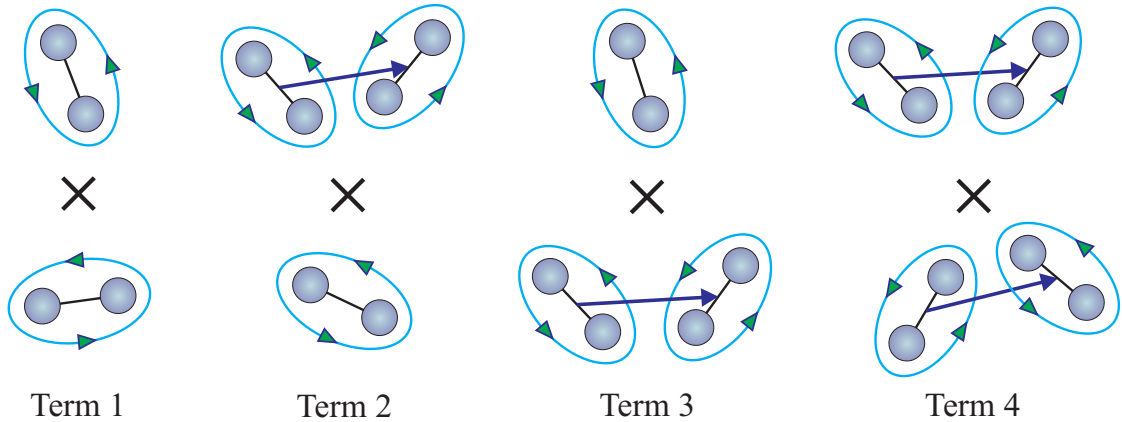}
\caption{\label{Fig:CorrContrib1}(Color online) Schematic illustration of different types of correlations contributing to the Fourier components $C_{q_{1},q_{2}}^{n}$ of the angular CCF corresponding to four different terms in Eq.~\ref{2.20a} (see text).}
\end{figure*}

In the case of a dense system, when the average distance $D$ between clusters is of the order of the size $d$ of a single cluster, the second sum in Eq.~(\ref{2.201}) can not be neglected. It can significantly affect the spectrum of
the angular CCF.
Taking both terms of Eq.~(\ref{2.201}) into account, the Fourier coefficients of the angular CCF can be written as the following sum of four terms
\begin{eqnarray}
C_{q_{1},q_{2}}^{n}&\propto& S^{n}_{1}+S^{n}_{2}+S^{n}_{3}+S^{n}_{4} \nonumber\\
&=&
\underbrace{\sum\limits _{k_{1}=k_{2}}^{N}\sum\limits _{k_{3}=k_{4}}^{N}...}_{\text{Term 1}}+
\underbrace{\sum\limits _{k_{1}=k_{2}}^{N}\sum\limits _{k_{3}\neq k_{4}}^{N}...}_{\text{Term 2}}+
\underbrace{\sum\limits _{k_{1}\neq k_{2}}^{N}\sum\limits _{k_{3}=k_{4}}^{N}...}_{\text{Term 3}}+
\underbrace{\sum\limits _{k_{1}\neq k_{2}}^{N}\sum\limits _{k_{3}\neq k_{4}}^{N}...}_{\text{Term 4}}
.\label{2.20a}
\end{eqnarray}
A schematic illustration of correlations corresponding to these four terms is shown in Fig.~\ref{Fig:CorrContrib1}.
The first term is a product of two 2-point correlation functions, each of which correlates a separate LS with itself and, therefore, contains the information only on the internal structure of LS's. This term does not depend on the density of a disordered system and was considered in detail in the previous subsection. The second and the third terms are the products of two 2-point correlation functions, one of which depends only on the internal structure of a LS, and another one correlates different LS's, separated in space. The fourth term is a product of two 2-point correlation functions each of which defines correlations of
spatially separated LS's. Clearly, the terms $S^{n}_{2}, S^{n}_{3}$ and $S^{n}_{4}$ depend on the density of a disordered system. In the following, we consider in detail each term of Eq.~(\ref{2.20a}).

Taking into consideration the results obtained in the previous subsection, the contribution of $S^{n}_{2}$ and $S^{n}_{3}$ in Eq.~(\ref{2.20a}) can be written as
\begin{eqnarray}
S^{n}_{2}+S^{n}_{3}=N\left[\langle e^{in\phi} \rangle L^{n\ast}({ q}_{1})
\sum\limits _{k_{1}\neq k_{2}}^{N}
L^{n}_{k_{1},k_{2}}({q}_{2})+
\langle e^{-in\phi} \rangle
\sum\limits _{k_{3}\neq k_{4}}^{N}
L^{n \ast}_{k_{3},k_{4}}({ q}_{1})
L^{n}({ q}_{2})\right],
\label{2.29}
\end{eqnarray}
where $L^{n}({q}_{j})$
and $L^{n}_{k_{1},k_{2}}({q}_{j})$ are defined in Eqs.~(\ref{LocStr1}, \ref{2.28a}), and the angular average
$\langle e^{in\phi} \rangle=(1/N)\sum\limits _{k=1}^{N} e^{in\phi_{k}}$ is defined similar to Eq.~(\ref{2.24}).
According to its structure the non-zero contributions of these terms are defined by the same selection rules
($L^{n}({q}_{j})\neq 0$) as for the term $S^{n}_{1}$. In addition, the value of non-zero contributions will be
modulated by the orientational order parameter $\langle e^{in\phi} \rangle$ and the spatial correlations between
different LS's defined by $L^{n}_{k_{1},k_{2}}({q}_{j})$.

Finally, for the fourth term in Eq.~(\ref{2.20a}) we have
\begin{eqnarray}
S^{n}_{4}=
\sum\limits _{k_{1}\neq k_{2}}^{N}
\sum\limits _{k_{3}\neq k_{4}}^{N}
L^{n\ast}_{k_{1},k_{2}}({q}_{1})
L^{n}_{k_{3},k_{4}}({q}_{2}),
\label{2.30}
\end{eqnarray}
where $L^{n}_{k_{1},k_{2}}({q}_{j})$ are defined in
Eq.~(\ref{2.28a}).
This term is determined only by the spatial correlations between different LS's,
and can play a significant role in the close-packed systems.

\bigskip

\begin{figure*}[!htbp]
\centering
\includegraphics[width=0.5\textwidth]{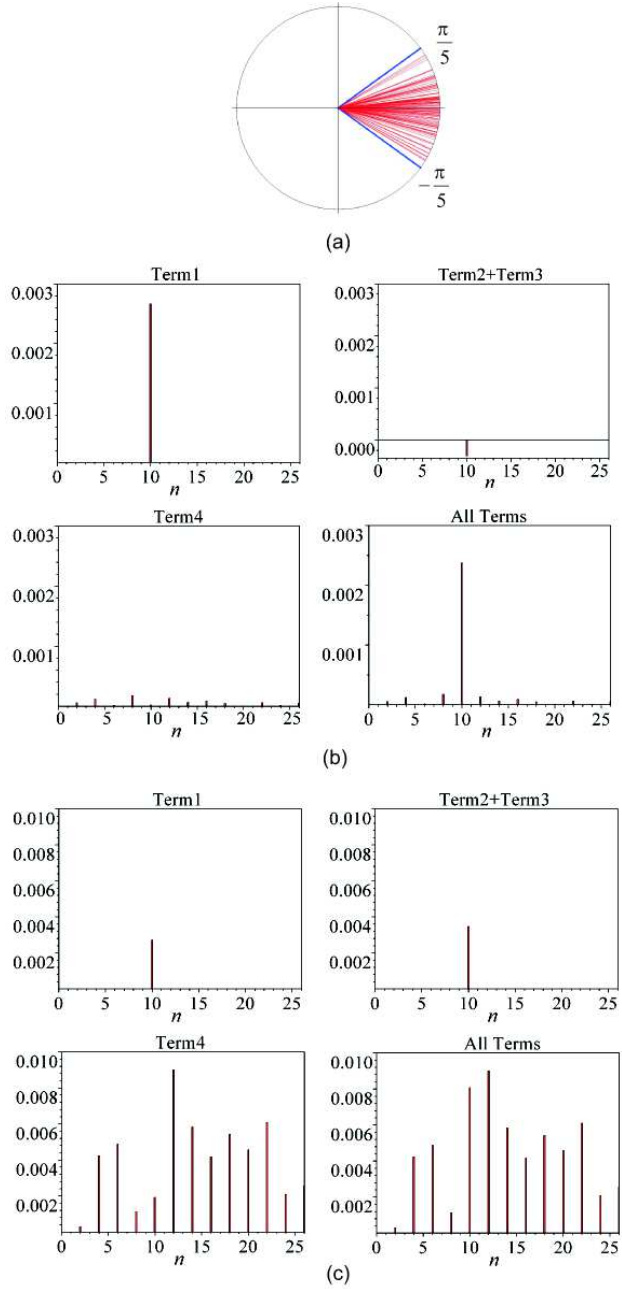}
\caption{\label{Fig:DiffTermContrib}(Color online) Normalized contributions of different terms to the Fourier
coefficients $C_{q}^{n}$ at $q=0.037\;\text{nm}^{-1}$. A 2D disordered system consisting of pentagonal clusters was considered. (a) Gaussian distribution of the in-plane angular orientations of the pentagonal clusters (with the standard deviation $\sigma=0.2*2\pi/5$). The blue lines bound a central angle $\phi=2\pi/5$. (b) The case of a dilute system ($D/d=150$).
(c) The case of a close-packed system ($D/d=1.5$).
}
\end{figure*}

We support our discussion by the calculations of
the normalized contributions of all four terms in Eq.~(\ref{2.20a}) to the Fourier coefficients $C_{q}^{n}$ (for the case $q_{1}=q_{2}=q$).
In our simulations we consider a 2D disordered system
consisting of pentagonal clusters [see Appendix~C and Fig.~\ref{Fig:SimpleShapes}(d)].
A high density system with $D/d=1.5$ and a low density system with $D/d=150$
were considered in these calculations, with the size of a pentagonal cluster $d=440\;nm$.
Both systems contain $121$ pentagonal clusters, and were characterized
by the same set of in-plane angular cluster orientations $\{\phi_{k}\}$. These angles were defined by the Gaussian distribution (\ref{GaussDistrFunc}), with the standard deviation $\sigma=0.2\cdot2\pi/5$ (see Fig.~\ref{Fig:DiffTermContrib}(a)). This distribution of angles covers all possible orientations for
a $5$-fold pentagonal cluster.

The results of the calculations for $q=0.037\;\text{nm}^{-1}$ are shown in Fig.~\ref{Fig:DiffTermContrib}(b, c) (here, for clarity, only the first 25 Fourier coefficients are shown).
In the case of a dilute system [Fig.~\ref{Fig:DiffTermContrib}(b)], the contribution of the Term 1 is strongly dominating over the contributions of all other terms. It reveals 5-fold symmetry by the presence of
the Fourier coefficient with $n=10$. Higher orders ($n=20,30,...$) are not present due to the choice of the $q$ value.
For this system the contribution from spatial correlations between different structures is negligible.
In the opposite case of the close-packed system [Fig.~\ref{Fig:DiffTermContrib}(c)], the
contribution from spatial correlations (Term 4) dominates over the contribution from the local symmetry of individual clusters (Term 1). The fourth term significantly modifies the frequency spectra, in particular by adding the
coefficients which are not related to the internal structure of clusters.
In this case, the Fourier coefficients with $n=10$ and $n=12$ are dominant in the shown range of the $C_{q}^{n}$ spectrum, but only one of them, with $n=10$, is related to the internal structure of clusters.

Our results show, that for dilute disordered systems
the main contribution to the cross-correlation
function $C_{q_{1},q_{2}}(\Delta) $ is defined by the local symmetry of clusters.
For a partially ordered system, one can extract this information by analyzing Fourier coefficients of the CCF.
For a dense system, the spatial correlations between clusters can become dominant, and their contribution to the CCF can not be easily separated from the contribution defined by the internal structure of clusters forming the system.


\section{Correlations in 3D systems. Ewald sphere curvature effects}


In our previous discussion of scattering on 2D systems, we have seen that only \textit{even} Fourier coefficients of the CCF have non-zero values. Here we will show, that non-zero \textit{odd} Fourier coefficients can be present due
to scattering to high angles on 3D systems due to Ewald sphere curvature effects.
In this case full expressions [Eqs.~(\ref{Eq:In_1}, \ref{Eq:Ln_k1k2_1})] containing $z$-components of the scattering vector $q^{z}_{j}$ have to be analyzed.

To simplify our discussion, we will consider here a 3D system consisting of identical 3D
clusters composed of identical point scatterers. The modified electron density (\ref{Eq:Ro_2}) of a cluster can be defined in the following form
\begin{equation}
\widetilde{\rho}_{k}(\mathbf{ r}^{\perp},q^{z}_{j})=f(q_{j})
\sum\limits _{i=1}^{N_{s}}\delta(\mathbf{ r}^{\perp}-\mathbf{ r}^{\perp}_{i})e^{-iq^{z}_{j}z_{i}},
\label{Eq:Ro_4}
\end{equation}
where $f(q_{j})$ is a form-factor of a scatterer, and $N_{s}$ is a number of scatterers in the cluster. The coordinates $(\mathbf{ r}^{\perp}_{i},z_{i})$ define the position of the $i$-th scatterer inside the cluster $k$.
Performing the integration in Eq.~(\ref{Eq:Ln_k1k2_1}) gives

\begin{eqnarray}
{ L}_{k_{1},k_{2}}^{n}({q}_{j}^{\perp} ,q_{j}^{z}) =\left|f(q_{j})\right|^{2}
\sum\limits _{l,m=1}^{N_{s}}
 e^{-iq^{z}_{j}z_{ml}}
J_{n}({ q}_{j}^{\perp}|\mathbf{ R}_{k_{2},k_{1}}^{\perp ml}|)e^{-in\phi_{\mathbf{ R}_{k_{2},k_{1}}^{\perp ml}}},
\label{Eq:Ln_k1k2_2}
\end{eqnarray}
where the summation over index $l$ is performed over the positions of scatterers in the cluster $k_{1}$, and
 the summation over index  $m$ is performed over the positions of scatterers in the cluster $k_{2}$.
 Substituting this expression into Eq.~(\ref{Eq:In_1}) we obtain
 \begin{eqnarray}
{ I}^{n}({q}_{j}^{\perp} ,q_{j}^{z}) =(i)^{n}\left|f(q_{j})\right|^{2}
\sum\limits _{k_{1},k_{2}=1}^{N}
\sum\limits _{l,m=1}^{N_{s}}
 e^{-iq^{z}_{j}Z_{k_{2},k_{1}}^{ml}}
J_{n}({ q}_{j}^{\perp}|\mathbf{ R}_{k_{2},k_{1}}^{\perp ml}|)e^{-in\phi_{\mathbf{ R}_{k_{2},k_{1}}^{\perp ml}}}.
\label{Eq:In_k1k2_3}
\end{eqnarray}
We note here that for $n\neq 0$ the terms with  $k_{1}=k_{2}$ and $l=m$ are equal to zero.
Taking now into account
that the terms with interchanged indices, i.e. $k_{1}, k_{2}$ and $k_{2}, k_{1}$, as well as $l, m$ and $m, l$,
differ from each other by a change of the sign of $Z_{k_{2},k_{1}}^{ml}$ and by an additional factor $(-1)^{n}$, which arises due to the change of the phase $\phi_{\mathbf{ R}_{k_{2},k_{1}}^{\perp ml}}=\phi_{\mathbf{ R}_{k_{1},k_{2}}^{\perp lm}}+\pi$,
we have for $\textit{even}$ values of $n$ in Eq.~(\ref{Eq:In_k1k2_3})
\begin{eqnarray}
{I}^{n}({ q}_{j}^{\perp} ,q_{j}^{z}) =2(i)^{n} \left|f(q_{j})\right|^{2} \sum\limits_{\begin{subarray}{c}
1\leq k_{1} \leq N\\
k_{1}\leq k_{2} \leq N
\end{subarray}}
\sum\limits_{\begin{subarray}{c}
1\leq l \leq N_{s}\\
l\leq m \leq N_{s}
\end{subarray}}
\cos{(q^{z}_{j}Z_{k_{2},k_{1}}^{ml})}
J_{n}({ q}_{j}^{\perp}|\mathbf{ R}_{k_{2},k_{1}}^{\perp ml}|)e^{-in\phi_{\mathbf{ R}_{k_{2},k_{1}}^{\perp ml}}},
\label{Eq:Ln_k1k2_4}
\end{eqnarray}
and for $\textit{odd}$ values of $n$:

\begin{eqnarray}
{I}^{n}({ q}_{j}^{\perp} ,q_{j}^{z}) =2(i)^{n+1} \left|f(q_{j})\right|^{2} \sum\limits_{\begin{subarray}{c}
1\leq k_{1} \leq N\\
k_{1}\leq k_{2} \leq N
\end{subarray}}
\sum\limits_{\begin{subarray}{c}
1\leq l \leq N_{s}\\
l\leq m \leq N_{s}
\end{subarray}}
\sin{(q^{z}_{j}Z_{k_{2},k_{1}}^{ml})}
J_{n}({ q}_{j}^{\perp}|\mathbf{ R}_{k_{2},k_{1}}^{\perp ml}|)e^{-in\phi_{\mathbf{ R}_{k_{2},k_{1}}^{\perp ml}}}.
\label{Eq:Ln_k1k2_3}
\end{eqnarray}

From the performed analysis we can see that, due to the curvature of the Ewald sphere (non-zero $q^{z}_{j}$ component),
we obtain non-zero odd Fourier components of CCF in scattering on a 3D system. These components become negligibly small at experimental conditions corresponding to the flat Ewald sphere, considered in the previous section.
A detailed discussion of differences between correlation analysis of 2D and 3D systems, based on simulations, will be given in the forthcoming paper.


\section{Conclusions and Outlook}


The basic results of this paper, Eq.~(\ref{Eq:In_1}) and (\ref{Eq:Ln_k1k2_1}), are characterized by the following structure.

1. They break up as a sum over LS pairs. Two points belonging to two LS's of a pair define a phase factor through the angle of the projection of their connecting vector on the $(x,y)$ plane.

2. Additional oscillating factors come from the the Bessel functions of integer order depending on the projections on the $(x,y)$ plane of the connecting vector and of the scattering vector $\mathbf{ q}_{j}^{\perp}$; and also from the effective density $\widetilde{\rho}_{k}(\mathbf{ r}^{\perp},q_{j}^{z})$. Note that, in the far-field diffraction limit adopted here, for scattering at small angles (small $q_{z}$), odd $m$ values are strongly suppressed in comparison to even ones by the trigonometric pre-factors. This is in disagreement with the strong $m=5$ components observed in the experiment \cite{PNAS}, and will need an additional analysis, for example, in the near-field scattering geometry.

3. Classes of LS pairs for which the oscillating factors systematically have the same sign give the largest contribution to the sum for a given $n$. This is the case for the $k_1 = k_2$ pairs, and the (purely two-dimensional) examples described in Appendix~C show how the non-vanishing values of $n$ are related to the rotational symmetry of the LS's around a common axis aligned with the direction of incidence. However, each LS contributes a value multiplied by a phase factor related to its orientation with respect to a reference direction in the plane; it is then easy to see that if the ensemble of illuminated LS's has a completely random orientation around the n-fold axis, the sum vanishes: in this case, indeed, LS's rotated by $\pi/n$ with respect to a given direction are as probable as those lined up in that direction, and their respective contributions cancel in the total result. This is in agreement with the concept of \textit{bond orientational order} \cite{Bru}. If a non-zero Fourier coefficient is observed, it implies either a preferential alignment along a given direction, either by a specific physical reason or, alternatively, because the ensemble of probed LS's is small enough to display pronounced fluctuations from the average uniform distribution of orientations; another interesting possibility, in view of the imminent availability of free-electron laser sources, could occur if the acquisition time is short enough to provide an "instantaneous" view, without effectively performing a time-average that necessarily restores the equal probability of all orientations. This may indeed be already the case in experiments involving very slow dynamics, as may be the case in \cite{PNAS}.

4. In a three-dimensional fluid, the order parameter defined above is contributed only by those molecules for which the $n$-fold axis is, at least to some degree of approximation, aligned to the direction of incidence. This probably explains why the observed Fourier components, especially in the intensity, but also in the CCF, are weak when compared to the extremely marked ones observed in hexatic liquid crystals, which are stacks of two-dimensional manifolds \cite{Pindak,Gorecka,Chou}. It is tempting to speculate that the subset of LS's with an approximate line-up of a symmetry axis, in a three-dimensional system, is always "dilute", in the sense that it is constituted by a small fraction of the total number of molecules or clusters. This would allow the application of results obtained in this paper for the dilute limit also to systems which are, in the three-dimensional sense, close-packed. In the companion paper, simulations are performed also with the purpose of establishing the extent of the deviation from perfect alignment of the symmetry axis which is compatible with an observable contribution to the CCF signal. It is important to bridge the gap between a two-dimensional theoretical interpretation that seems to arise naturally from the experimental geometry and the three-dimensional isotropy of ordinary samples.

There are various directions that future experiments may explore; it would certainly be very interesting to monitor the CCF signal in a system in which a controllable experimental parameter (e.g. temperature, an electric or magnetic field) may provide a way to vary the degree of alignment of a symmetry axis; or in which the bond orientational order is well characterized.

\bibliography{XCCA_theory_paper_literature}

\appendix

\section{} 


Here we calculate the Fourier coefficients
\begin{equation}
I^{n}({q}^{\perp},q^{z})=\frac{1}{2\pi}\int_{0}^{2\pi}I(\mathbf{q})e^{-in\varphi}d\varphi
\label{Eq:App2}
\end{equation}
of the intensity scattered at certain momentum transfer vector $\mathbf{ q}$, defined in Eq.~(\ref{Intens2}). The scalar product $\mathbf{ q}^{\perp}\cdot \mathbf{ R}_{k_{2},k_{1}}^{\perp 21}$ in the exponent of (\ref{Intens2}) can be written as
\begin{eqnarray}
\mathbf{ q}^{\perp}\cdot \mathbf{ R}_{k_{2},k_{1}}^{\perp 21} & = & { q}^{\perp}\cdot|\mathbf{ R}_{k_{2},k_{1}}^{\perp 21}|\cos(\varphi-\phi_{\mathbf{ R}_{k_{2},k_{1}}^{\perp 21}}),
\label{Eq:App3}
\end{eqnarray}
where $q^{\perp},\varphi$
are the polar coordinates of the perpendicular component of the vector
$\mathbf{ q}^{\perp}$ [see Fig.~\ref{Fig:EwaldSphere}(a)] and $|\mathbf{ R}_{k_{2},k_{1}}^{\perp 21}|$, $\phi_{\mathbf{ R}_{k_{2},k_{1}}^{\perp 21}}$
are the polar coordinates of the perpendicular components of the vector
$\mathbf{ R}_{k_{2},k_{1}}^{\perp 21}$ (see Fig.~\ref{Fig:Sample}).
Substituting this expression in Eq.~(\ref{Intens2}) and using the Jacobi-–Anger expansion\cite{Cuyt} of the exponential
functions in series of Bessel functions $J_{n}(\rho)$ of the first kind of integer order $m$
\begin{equation}
e^{i\rho\cos\varphi}=\sum\limits _{m=-\infty}^{\infty}(i)^{m}J_{m}(\rho)e^{im\varphi} \nonumber
\end{equation}
we can write
\begin{eqnarray}
I^{n}({q}^{\perp},q^{z})&=&\sum\limits _{k_{1},k_{2}=1}^{N}e^{-iq^{z}\cdot Z_{k_{2},k_{1}}}
 \int\int\widetilde{\rho}_{k_{1}}^{\ast}(\mathbf{ r}_{1}^{\perp},q^{z})\widetilde{\rho}_{k_{2}}(\mathbf{ r}_{2}^{\perp},q^{z})
 d\mathbf{ r}^{\perp}_{1}d\mathbf{ r}^{\perp}_{2}\times \nonumber\\
 &&\int_{0}^{2\pi}\frac{d\varphi}{2\pi}\sum\limits _{m=-\infty}^{\infty}
 (i)^{m}J_{m}({ q}^{\perp}|\mathbf{ R}_{k_{2},k_{1}}^{\perp 21}|)
 e^{-im\phi_{\mathbf{ R}_{k_{2},k_{1}}^{\perp 21}}}
 e^{i(m-n)\varphi}.
\label{Eq:App4}
\end{eqnarray}
Integration over $\varphi$ in (\ref{Eq:App4}) gives
\begin{equation}
\int\limits _{0}^{2\pi}(d\varphi/2\pi)\exp[i(m-n)\varphi]=\delta_{m,n}, \nonumber
\end{equation}
where $\delta_{m,n}$ is the Kroneker symbol. Substitution the
result of this integration into (\ref{Eq:App4}) finally gives
\begin{eqnarray}
I^{n}({q}^{\perp},q^{z})&=&(i)^{n}\sum\limits _{k_{1},k_{2}=1}^{N}e^{-iq^{z}\cdot Z_{k_{2},k_{1}}}
 \int\int\widetilde{\rho}_{k_{1}}^{\ast}(\mathbf{ r}_{1}^{\perp},q^{z})\widetilde{\rho}_{k_{2}}(\mathbf{ r}_{2}^{\perp},q^{z})
 d\mathbf{ r}^{\perp}_{1}d\mathbf{ r}^{\perp}_{2}\times \nonumber\\
 &&J_{n}({q}^{\perp}|\mathbf{ R}_{k_{2},k_{1}}^{\perp 21}|)
 e^{-in\phi_{\mathbf{ R}_{k_{2},k_{1}}^{\perp 21}}}.
\label{Eq:App5}
\end{eqnarray}%

It is clear from the definition (\ref{Eq:App2}) that
\begin{equation}
\left\langle I(\mathbf{ q},\varphi)\right\rangle _{\varphi}=I^{0}({q}^{\perp},q^{z})=\sum\limits _{k_{1},k_{2}=1}^{N}e^{-iq^{z}\cdot Z_{k_{2},k_{1}}}
\int\int d\mathbf{ r}_{1}^{\perp}d\mathbf{ r}_{2}^{\perp}\widetilde{\rho}_{k_{1}}^{\ast}(\mathbf{ r}_{1}^{\perp},q^{z})\widetilde{\rho}_{k_{2}}(\mathbf{ r}_{2}^{\perp},q^{z})J_{0}({q}^{\perp}|\mathbf{ R}_{k_{2},k_{1}}^{\perp 21}|).
\label{Eq:App10}
\end{equation}
These results imply that the Fourier coefficients $D_{I}^{n}(q)$ of the normalized deviation, defined in
(\ref{Eq:Dq_1}), can be written as follows
\begin{equation}
D_{I}^{n}( q)=\left\{ \begin{array} {ll}
(I^{n}({ q}^{\perp},q^{z}))/(I^{0}({ q}^{\perp},q^{z}))& \text{ if }n \neq 0\\
0& \text{ if }n=0.
\end{array} \right.
\label{Eq:App11}
\end{equation}


\section{} 


Here, we will prove, that for dilute systems the main contribution to the Fourier coefficients of intensity ${I}^{n}({q}_{j})$ is given by the first term in the expansion (\ref{2.201}).
In the case of a dilute disordered system, when typical distances between
LS's are much larger than the size of a LS's itself, i.e.,
$|\mathbf{ R}_{k_{l},k_{m}}^{ij}|\gg|\mathbf{ r}_{ij}|$, we can use the following approximation for the values of the Bessel functions in the integrals ${ L}_{k_{1},k_{2}}^{n}({ q}_{j} ,q_{j}^{z})$ in Eq.(\ref{Eq:Ln_k1k2_1}):
$
J_{n}( { q}_{j}|\mathbf{ R}_{k_{2},k_{1}}^{21}|)=
J_{n}({ q}_{j}|\mathbf{ R}_{k_{2},k_{1}} +\mathbf{ r} _{21}|)
\simeq  J_{n}({q}_{j}|\mathbf{ R}_{k_{2},k_{1}} |).
\label{2.27}
$
Furthermore, for large values of the argument of the Bessel function $ J_{n}( \rho)$,
$
\rho\gg \left(  {n^2}/{2}-{1}/{8} \right ),
$
one can use the asymptotic expansion\cite{{Abramowitz}}\linebreak
$
J_{n}( \rho)\simeq
\sqrt{\frac{2}{\pi\rho}}
\cos\left(\ \rho -\frac{n\pi}{2}-\frac{\pi}{4} \right).
$
Taking all this into account we, finally, get for the integral $L_{k_{1}\neq k_{2}}^{n}$ in Eq.~(\ref{Eq:Ln_k1k2_1}) in the case of a 2D system
\begin{eqnarray}
L_{k_{1}\neq k_{2}}^{n}({q}_{j}) & = &
\int\int d\mathbf{ r}_{1} d\mathbf{ r}_{2} \widetilde{\rho}_{k_{1}}^{\ast}(\mathbf{ r}_{1} )\widetilde{\rho}_{k_{2}}
(\mathbf{ r}_{2} )J_{n}({q}_{j}|\mathbf{ R}_{k_{2},k_{1}}^{21}|)e^{-in\phi_{\mathbf{ R}_{k_{2},k_{1}}^{21}}}\label{2.28a}\\
& \simeq &
\sqrt{\frac{2}{\pi{ q}_{j}|\mathbf{ R}_{k_{2},k_{1}} |}}
\cos\left(\ { q}_{j}|\mathbf{ R}_{k_{2},k_{1}} | -\frac{n\pi}{2}-\frac{\pi}{4} \right)
e^{in\phi_{\mathbf{ R}_{k_{2},k_{1}} }}   P^{*}_{k_{1}} P_{k_{2}},
\label{2.28}
\end{eqnarray}
where
$$
P_{k} = \int d\mathbf{ r}  \widetilde{\rho}_{k}(\mathbf{ r} ).
$$
In deriving Eq.~(\ref{2.28}) we also used an approximation
$\phi_{\mathbf{ R}_{k_{2},k_{1}}^{\perp 21}}\approx \phi_{\mathbf{ R}_{k_{2},k_{1}} }$
that is valid for dilute systems and means that this phase does not depend on individial orientations of LS's, but is determined by their positions in the system.

We analyze now this asymptotic behavior of the second sum in Eq.~(\ref{2.201}). According to Eq.~(\ref{2.28}),
for the given $q_j$ value the function $L_{k_{1}\neq k_{2}}^{n}({q}_{j})$ decays as
$1/\sqrt{{ q}_{j}|\mathbf{ R}_{k_{2},k_{1}} |}$ with the increase of the distance $\mathbf{ R}_{k_{2},k_{1}}$ between the LS's.
At the same time, it can be noted that the sum $\sum\limits _{k_{1}\neq k_{2}}^{N}...$ in Eq.~(\ref{2.201})
contains $N(N-1)$ terms comparing to $N$ terms in the sum $\sum\limits _{k_{1}=k_{2}=k}^{N}...$. However, the presence in Eq.~(\ref{2.28}) of the exponential factor $e^{in\phi_{\mathbf{ R}_{k_{2},k_{1}} }}$ with random phases $\phi_{\mathbf{ R}_{k_{2},k_{1}} }$, corresponding to a big number of LS's present in dilute system, will additionally reduce the contribution of the second sum in Eq.~(\ref{2.201}).
Therefore, for dilute systems the dominant contribution to the Fourier coefficients $C_{q_{1},q_{2}}^{n}$ will be defined by the first sum in Eq.~(\ref{2.201}) with $k_{1}=k_{2}$.


\section{} 

\begin{figure*}[!htbp]
\centering
\includegraphics[width=0.6\textwidth]{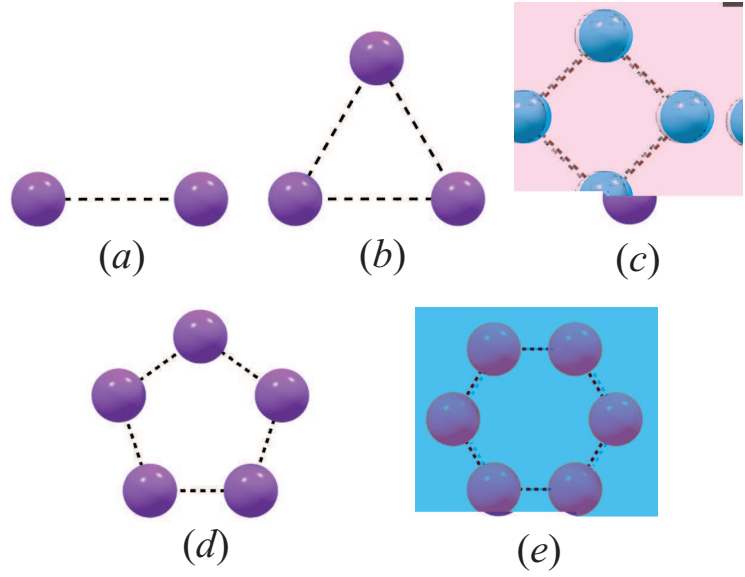}
\caption{\label{Fig:SimpleShapes}(Color online) 2D particles with simple geometrical shapes owning different
rotational symmetries (rotational axes are perpendicular to the plane of the figure): (a) 2-fold, (b) 3-fold, (c) 4-fold, (d) 5-fold and (e) 6-fold.}
\end{figure*}

We consider here simple 2D structures (clusters) with distinct
rotational $m$-fold symmetries shown in Fig.~\ref{Fig:SimpleShapes}.
We define the electron density of a cluster as a real-valued quantity in the following form
\begin{equation}
\rho(\mathbf{ r}_{l})=
\sum\limits _{i=1}^{N_{s}}f_{i}(q_{j})\delta(\mathbf{ r}-\mathbf{ r}_{i})=
\sum\limits _{i=1}^{N_{s}}f_{i}(q_{j})\delta(r_{l}-r_{i})\delta(\phi_{l}-\phi_{i}),
\label{AIII.1}
\end{equation}
where $N_{s}$ is a number of scatterers in the cluster, $f_{i}(q_{j})$ is a scattering factor of the $i$-th scatterer in the cluster, $\mathbf{ r}=(r,\phi)$,
$r$ and $\phi$ are the polar coordinates of a scatterer in the cluster.
Using the definition (\ref{AIII.1}) and performing the integration in Eq.~(\ref{LocStr1}) we get
\begin{eqnarray}
&&L^{n}({q}_{j})=\sum\limits _{s,t=1}^{N_{s}}f^{\ast}_{s}(q_{j})f_{t}(q_{j})J_{n}({q}_{j}
|\mathbf{ r}_{ts} |)e^{-in\phi_{\mathbf{ r}_{ts} }}.\label{LocStr3}
\end{eqnarray}

For an arbitrary cluster
with $m$-fold rotational symmetry shown in Fig.~\ref{Fig:SimpleShapes} the following assumptions are valid:
$r_{i}=a$, i.e., all scatterers in the cluster are
located on equal distances from its center, $\phi_{i}=2\pi/m\cdot(i-1),\:i=1... N_{s}$, where $m=N_{s}$ is a highest order of rotational symmetry in the cluster, and we also assume $f_{i}(q_{j})=f(q_{j})$.
Using these assumptions in Eq.~(\ref{LocStr3}), we derive the expressions of $L^{n}({ q}_{j})$  for each of the clusters shown in Fig.~\ref{Fig:SimpleShapes}.

\begin{enumerate}
\item $m=2$, Fig.~\ref{Fig:SimpleShapes}(a):
\begin{equation}
L^{n}({ q}_{j})=
\left\{
\begin{array}{l l}
2|f(q_{j})|^{2}[J_{n}(0)+J_{n}(2a{ q}_{j})]& \text{if }n \text{ mod } 2=0\\
0 & \text{if }n \text{ mod } 2\neq 0
\end{array} \right.
\label{AIII.2fold}
\end{equation}
\item $m=3$, Fig.~\ref{Fig:SimpleShapes}(b):
\begin{equation}
L^{n}({ q}_{j})=
\left\{
\begin{array}{l l}
3|f(q_{j})|^{2}[J_{n}(0)+2J_{n}(\sqrt{3}a{ q}_{j})]&  \text{if }n \text{ mod } 12=0\\
-6|f(q_{j})|^{2}J_{n}(\sqrt{3}a{ q}_{j})& \text{if }n \text{ mod } 6=0,\;n \text{ mod } 12\neq0\\
0 & \text{other } n
\end{array} \right.
\label{AIII.3fold}
\end{equation}
\item $m=4$, Fig.~\ref{Fig:SimpleShapes}(c):
\begin{equation}
L^{n}({ q}_{j})=
\left\{
\begin{array}{l l}
4|f(q_{j})|^{2}\{J_{n}(0)+J_{n}(2a{ q}_{j})+2J_{n}(\sqrt{2}a{ q}_{j})\}&\text{if }n \text{ mod } 8=0\\
4|f(q_{j})|^{2}[J_{n}(2a{ q}_{j})-2J_{n}(\sqrt{2}a{ q}_{j})]& \text{if }n \text{ mod } 4=0,\\
&\quad n \text{ mod } 8\neq0\\
0 & \text{other } n
\end{array} \right.
\label{AIII.4fold}
\end{equation}
\item $m=5$, Fig.~\ref{Fig:SimpleShapes}(d):
\begin{equation}
L^{n}({ q}_{j})=
\left\{
\begin{array}{l l}
5|f(q_{j})|^{2}\{J_{n}(0)+2[J_{n}(A_{1}a{ q}_{j})+J_{n}(A_{2}a{ q}_{j})]\}& \text{if }n \text{ mod } 20=0\\
-10|f(q_{j})|^{2}[J_{n}(A_{1}a{ q}_{j})+J_{n}(A_{2}a{ q}_{j})]& \text{if }n \text{ mod } 10=0,\\
&\quad n \text{ mod } 20\neq0\\
0 & \text{other } n,
\end{array} \right.
\label{AIII.5fold}
\end{equation}
where $A_{1}=\sqrt{\frac{1}{2}(5-\sqrt{5})}$, $A_{2}=\sqrt{\frac{1}{2}(5+\sqrt{5})}$.

\item $m=6$, Fig.~\ref{Fig:SimpleShapes}(e):
\begin{equation}
L^{n}({ q}_{j})=
\left\{
\begin{array}{l l}
6|f(q_{j})|^{2}\{J_{n}(0)+2J_{n}(a{ q}_{j})+J_{n}(2a{ q}_{j})+2J_{n}(\sqrt{3}a{ q}_{j})\}& \text{if }n \text{ mod } 12=0\\
6|f(q_{j})|^{2}[2J_{n}(a{ q}_{j})+J_{n}(2a{ q}_{j})-2J_{n}(\sqrt{3}a{ q}_{j})]& \text{if }n \text{ mod } 6=0,\\
&\quad n \text{ mod } 12\neq0\\
0 & \text{other } n
\end{array} \right.
\label{AIII.6fold}
\end{equation}

\end{enumerate}

Equations (\ref{AIII.2fold}-\ref{AIII.6fold}) define the selection rules which determine the
contributions to the $n$-th coefficient $C_{q_{1},q_{2}}^{n}$ related only to the internal structure of clusters.
For instance, Eq.~(\ref{AIII.3fold}) means that the contribution from the internal structure
of the cluster shown in Fig.~\ref{Fig:SimpleShapes}(b)
to the Fourier coefficients with $n=6,18,30,...$ is defined by the function
$L^{n}({ q}_{j})=-6|f(q_{j})|^{2}J_{n}(\sqrt{3}a{ q}_{j})$; for the coefficients
with $n=12,24,48,...$, $L^{n}({ q}_{j})=6|f(q_{j})|^{2}J_{n}(\sqrt{3}a{ q}_{j})$; for $n=0$ coefficient $L^{n}({ q}_{j})=3|f(q_{j})|^{2}[1+2J_{0}(\sqrt{3}a{ q}_{j})]$, while other Fourier coefficients do not contain
any information on the internal structure of this particular cluster.

\newpage

\end{document}